\documentclass[11pt]{article}
\newcommand{\comm}[1]{}

\usepackage{amssymb,epsfig}
\usepackage{amsthm}
\usepackage[numbers]{natbib}
\def\citet{\cite}
\usepackage{etoolbox}
  \renewcommand{\theequation}{\arabic{equation}}
\usepackage{array}
\usepackage{tabu}
\usepackage{hyperref}
\usepackage{graphicx}
\graphicspath{{C:/Users/valough/Desktop/MATH5001/}}
\usepackage{multirow}
\usepackage{tabularx}

\newcommand{\be}{\begin{equation}}
\newcommand{\ee}{\end{equation}}
\newcommand{\bd}{\begin{displaymath}}
\newcommand{\ed}{\end{displaymath}}
\newcommand{\ba}{\begin{array}{ll}}
\newcommand{\ea}{\end{array}}
\newcommand{\baa}{\begin{eqnarray}}
\newcommand{\eaa}{\end{eqnarray}}
\newcommand{\baaa}{\begin{eqnarray*}}
\newcommand{\eaaa}{\end{eqnarray*}}

\def\y{\mu}

\title{Some examples of application for predicting of compressive sensing method}
\author{
Nicholas James Rowe }
\begin{document}
\maketitle
\let\thefootnote\relax\footnote{School of Electrical Engineering, Computing and Mathematical Sciences, Curtin
University,   GPO Box U1987, Perth, 6845 Western Australia.}
\begin{abstract}
This paper considers application of the “SALSA” algorithm as a method of forecasting and applies it to simulated electrical signal, temperature recording from the Australian Bureau of Meteorology and stock prices from the Australian stock exchange. It compares it to basic linear extrapolation and casual smoothing extrapolation, in all cases SALSA extrapolation proves to be a better method of forecasting than linear extrapolation. However, it cannot be imperially stated that it is superior to Causal smoothing extrapolation in complex systems as it has a higher L2-euclidean in these experiments. while usually retaining more shape and statistical elements of the original function than Causal smoothing extrapolation. Leading to the conclusion the Causal Smoothing extrapolation can provide a more conservative forecast for complex systems while the SALSA algorithm more accurately predicts the range of possible events as well as being the superior forecasting method for electrical signals, the physical process it is designed to forecast.
\end{abstract}
\section{Introduction}
Forecasting future events has always been both an extremely worthwhile and difficult endeavour, and while many processes have been able to predicted by improving our fundamental understanding of the underlying process, not all processes are simple and stable enough to mathematically predicted. The purpose of this project is to do a comparative study on the accuracy of extrapolations with the algorithm known as “SALSA” and a previously explored method “Causal Smoothing Extrapolation”. Applying both to real world data and evaluating how accurately each of algorithms can predict data in comparison to each other and evaluating how useful these forecasts may be.
Causal Smoothing Extrapolation  proposed in “On Causal Extrapolation of sequences with application to forecasting” [1]  was the primary focus of the previous work [2], in which it was determined that in most cases more Causal Smoothing Extrapolation creates more accurate forecasts than linear extrapolation. This was explored using financial and regional maximum temperature time series data. The following next step in the investigation of Causal Extrapolation would be to compare it with a more complex extrapolation method. “Split augmented Lagrangian Shrinkage algorithm” or “Salsa” first proposed in “ Fast Image Recovery using Variable splitting and Constrained Optimization” [3] is a method of data extrapolation currently under study by todays mathematics community and is there for the most relevant method to compare to Causal Smoothing Extrapolation.
\break
\section{Theory}
\subsection{Causal Extrapolation Summary}
The method is based on approach from [1].  The following is an extract from “Applications of band-limited extrapolation to forecasting of weather and financial time series” [2] which provides a summary of how the Causal Smoothing Extrapolation works.
$$(Q^*z)_k=\frac{\Omega}{\pi} \sum_{t=q}^s sinc(k{\pi} + {\Omega}t)z(t)$$
$$R_{km}=(\frac{\Omega}{\pi})^2 \sum_{t=q}^s sinc(m{\pi} + {\Omega}t)sinc(k{\pi} + {\Omega}t)$$
$$R_v=R+vI$$
$$y_k = R_v^{-1}Q^*x$$
$$\widehat{x}(t)=Qy_k$$
$$\widehat{x}(t)=(Qy)(t)\frac{\Omega}{\pi} \sum_{t=q}^s sinc(k{\pi} + {\Omega}t)$$
This shows the progression of time series data z(t) for $s\le t\le q $ from raw data to Causally Smoothed $x(t)$  points by operators $Q^*$, $Rv$ and $Q$. Where$ \omega$ and $v$ are constants selected by the simulation of 111 data points repeated 10,000 times in order to select the constants which give the most accurate projections. This topic is fleshed out further in “Applications of band-limited extrapolation to forecasting of weather and financial time series” [2] while the full proofs and workings of this algorithm can be found in [1]. 
\subsection{Split Augment Lagrangian Shrinkage Algorithm or SALSA}
SALSA is a method of image restoration and reconstruction where the goal is to minimise the following function by spliting the two major components of the functions and minimising them independently.$$\min_{x,v\epsilon R^n}=\frac{1}{2} ||Ax-y||_2^2+{\omega} {\sigma}(v)$$
Here component one $f(x)=\frac{1}{2}||Ax-y||_2^2$ and component two $f(x)={\omega} {\sigma}(v)$.
The algorithm for forecasting a signal is done by comparing a signal y with length M to a bases vector x length N transformed by A which is a MxN matrix and minimise the difference between the two.
$$y=Ax$$
$$y=\begin{tabular}{|c|}y(0)\\y(1)\\...\\y(M-1)\end{tabular} \ ,\ x=\begin{tabular}{|c|}x(0)\\x(1)\\...\\x(N-1)\end{tabular}$$

With multiple repetition of the SALSA algorithm the bases vector x and transform A can completely replicate the original signal y, when values of y are unknown it can interpolate and extrapolate the missing values of the original signal. 
\newline
The complete algorithm is the following: 
\newline
1. Set k=0,choose${\mu} > 0 ,\ v_0\ and\ d_0$.\\*
2. Repeat point\\*
3. $x_{k+1}=argmin_x||Ax-y||_2^2+{\mu}||x-v_k-d_k||_2^2$\\*
4. $v_{k+1}=argmin_v {\omega} {\sigma}(v)+\frac{\mu}{2} ||x_{k+1}-v_k-d_k||_2^2$\\*
5. $d_{k+1} =d_k-(x_{k+1}-v_{k+1})$\\*
6$.k{\leftarrow} k+1$\\*
7. Continue until stopping criterion is satisfied.\\*
\newline
Here $||x||_2^2$ is defined by the $L_2$ norm
$$||x||_2^2 = \sum_{n=0}^{N-1} |x(n)|^2. $$
\newline
The algorithm  will initialise  with zero vectors $d$ and $v$ of  the same size as $x$. The parameter  ${\mu}$ is the penalty parameter which will be determined by simulation, ${\sigma}$ is the regulation parameter(P in the appendices code) which can be found mathematically as N the length of the basis vector and ${\omega}$(lambda in the appendices code) is the regularizer also known as amount the white noise found in the signal. Elements in the basis vector x are updated with each iteration for as many repetitions as required.
This can be summarised as the following, after choosing initial conditions ${\mu}$>0, $v_0$, and $d_0$ apply to the time series bases vector $x_k$ to get $x_{k+1}$ repeat until the values of the bases vector set can be transformed into a vector which closely resembles the original data set. 
For my purposes I have chosen to use the “Fast Fourier transform” in Matlab to convert the original signal y to basis set x and the “Inverse Fast Fourier transform” to convert it back. Circumventing the need to directly define the transform A. I have also used the function ‘soft’ from Ivan Selesnicks SALSA toolbox in order to calculate the argmin values needed in step 3 and 4 of the algorithm [4].
Finally, in order to forecast using SALSA part of the original signal is masked using matrix K. K is and MxM identity matrix with values missing to represent a patchy or incomplete signal, since I am only using SALSA for forecasting the matrix K will obscure the last 2-10 values of y. The vector Ky will be used in place y in the algorithm. 
$$Ky=Ax$$
For further information on this please look into the lecture notes provided by Ivan Selesnick [4].
\subsection{Linear Extrapolation}
Linear extrapolation has also been included in portions of this paper in order to provide a base line for how the previous two methods compare to a standard trend line. The following equation is used in all linear forecasts in this paper.\\*
Linear extrapolation for A historical Data points:
$$\widehat{x}(t)=\frac{(z(t_0)-z(t_0)-A)}{A}(t) + (z(t_0)-\frac{(z(t_0)-z(t_0-A)}{A})t_0 $$

\section{Monte Carlo Simulation for SALSA forecast}
Since d,v,$ {\sigma}$ have been defined previously this leaves values ${\mu} , {\omega} $ and the length of the bases vector N to be selected. In order to do this a Monte Carlo simulation was created to test the most effective values for these constants. This follows the same process used in the previous experiments with Casual extrapolation [1,2]. The Monte Carlo simulation is structured as follows, A(t) takes random values within a uniform distribution, ${ \omega}(t) $is standard gaussian white noise and ${\rho}$ is simply 1 [1]. This is to ensure that the process appears to be entirely random and cannot be forecasted by some other method.\\*

$$
z(t)=A(t)z(t-1)+{\gamma}(t),\quad t=-1,1,2,.....
$$
Here  A(t) takes random values within a uniform distribution, $\gamma(t)$ is the standard Gaussian white noise.

\begin{figure}[h!]
\centering
\includegraphics[width=\textwidth]{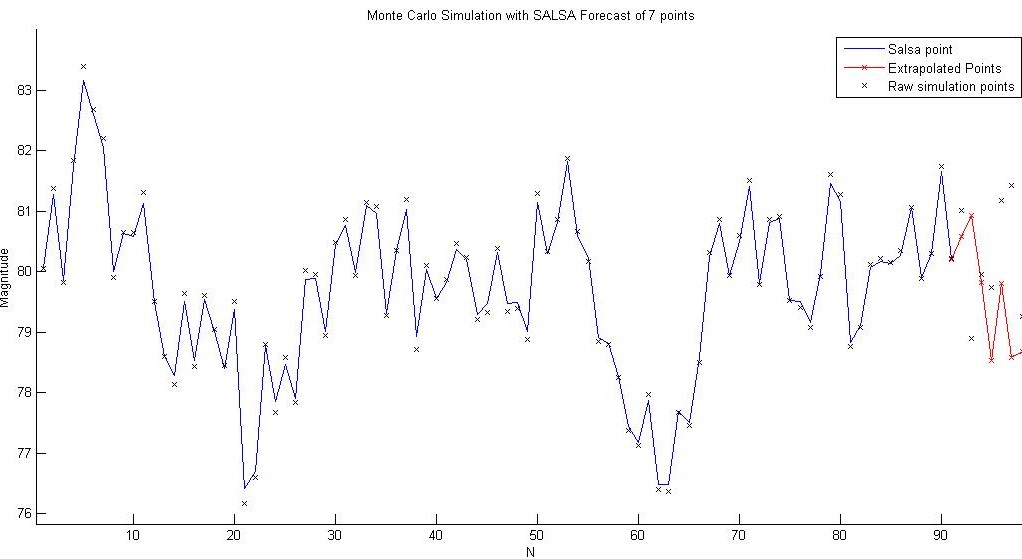}
\caption{Monte Carlo simulation, forecast of 10 point}
\end{figure}

We used $0.1\le \mu\le 200 $ in steps of 0.1, $1\le \omega\le 10$ with integer steps and $100\le N\le 1000$  in steps of 100 with 10,000 trials each forecasting 7 points for a total of 70,000 total data point per trial  was the original testing circumstances in order to provide a complete overview and find the most effective values, however this test was abandon after 72 hours of simulation when no significant reduction in the L2-Norm was noticed was noticed past ${\mu}$ of 0.6 , ${\omega}$ of 1 and N of 200.  This has henceforth been used at the as the values for future forecasting.
\begin{table}[h!]
\begin{tabular} { |c|c|c|c|c|c|c|c|c|c|c| }
 \hline
 ${\mu}$& 0.1&0.2&0.3&0.4&0.5&0.6&0.7&0.8&0.9&1.0 \\
 \hline
Residual per point &2.0767&2.1702&2.1743&2.1936&2.1788&2.1135&2.2526&2.1391&2.1597&2.1237 \\
\hline
 ${\mu}$&1.1&1.2&1.3&1.4&1.5	&1.6&1.7&1.8&1.9&2.0 \\
 \hline
Residual per point &2.2009&2.1778&2.2461&2.2261&2.1243&2.1720&2.1888&2.1460&2.1852&2.2436\\
\hline
\end{tabular}
\caption{Table of values of mean residual per point for values of ${\mu}$}
\end{table}

\section{Electrical signal experiements}
The electrical signal information is a simulated data set provided by Dr. Nikolai Dokuchaev to the author.
\index{
It consists of 1000 signals lasting for 3 minutes each. Two experiments were done using this data set, the first was a 0.2 second forecast repeated 450 times and 1 second forecast repeated 90 times. Both forecasting methods were allowed 91 points of data in order to predict the next 2-10 points. The Results of which are below, linear extrapolation has been intentionally omitted from these graphs in order to improve the clarity of the following figures.

\begin{table}[h!]
\begin{tabular}{|c|c|c|c||c|c|c|c|}
 \hline
 \multirow{2}{2em}{Data Type}& \multirow{2}{2em}{Raw Data}&\multirow{2}{6em}{Casual Extrapolation}&\multirow{2}{6em}{Salsa Extrapolation}&\multirow{2}{5em}{Comparison method}&\multirow{2}{5em}{Casual Forecast}&\multirow{2}{5em}{Salsa Forecast}&\multirow{2}{5em}{Linear Forecast} \\
&&&&&&&\\
 \hline
Min &-2.82&-2.814&-2.736&\multirow{2}{4em}{Total L2 residual}&\multirow{2}{4em}{41.044}&\multirow{2}{4em}{11.5482}&\multirow{2}{4em}{19.8732} \\
\cline{1-4}
 Max&2.684&2.682&2.558& & & &\\
\hline
Mean &-0.00531&3.08e-05&0.00112&\multirow{3}{4em}{Total L2 residual per point}&\multirow{3}{4em}{0.0456}&\multirow{3}{4em}{0.0128}&\multirow{3}{4em}{0.0221} \\
\cline{1-4}
STD &1.468&1.467&1.404&&&&\\
\cline{1-4}
Range &5.504&5.496&5.294&&&&\\
\hline
\end{tabular}
\caption{Statistical Results of Electrical signal 0.2 second forecast}
\end{table}

\begin{table}[h!]
\begin{tabular}{|c|c|c|c||c|c|c|c|}
 \hline
 \multirow{2}{2em}{Data Type}& \multirow{2}{2em}{Raw Data}&\multirow{2}{6em}{Casual Extrapolation}&\multirow{2}{6em}{Salsa Extrapolation}&\multirow{2}{5em}{Comparison method}&\multirow{2}{5em}{Casual Forecast}&\multirow{2}{5em}{Salsa Forecast}&\multirow{2}{5em}{Linear Forecast} \\
&&&&&&&\\
 \hline
Min &-2.82&-2.777&-2.658&\multirow{2}{4em}{Total L2 residual}&\multirow{2}{4em}{306.2743}&\multirow{2}{4em}{61.9608}&\multirow{2}{4em}{258.2352} \\
\cline{1-4}
 Max&2.684&2.672&2.542& & & &\\
\hline
Mean &-0.00531&-0.01101&-0.00812&\multirow{3}{4em}{Total L2 residual per point}&\multirow{3}{4em}{0.3437}&\multirow{3}{4em}{0.0695}&\multirow{3}{4em}{0.2898} \\
\cline{1-4}
STD &1.475&1.466&1.336&&&&\\
\cline{1-4}
Range &5.504&5.449&5.2&&&&\\
\hline
\end{tabular}
\caption{{Statistical Results of Electrical signal 1.0 second forecast}}
\end{table}
In both cases SALSA extrapolation has far lower L2 euclidean per point then linear extrapolation and casual smoothing extrapolation showing it to be far better extrapolation method for this process. This makes sense as the SALSA algorithim has been designed with the goals of processing and recovering signal data, it maintains a L2-euclidean of 0.0456 and 0.0695 per point for the 0.2 second and 1.0 second forecasts respectivly. This is 58\% of its nearest competitors residual in the 0.2 forecast and 24\% in the 1.0 second forecast. It also does not heavily compress the range of data points maintaining 94-96\% of the range of the raw data set . The inaccuracy of Casual smoothing in this case is highly visable on the 1 second forecast data, in which forecasts maintains a flat trajectory while the data sharply assends or decends. This highlights the need to forecast using methods that relate to the physical process whenever it is avalible and save Casual smoothing extrapolation for when the process is completely unknown.
\begin{figure}[h!]
\centering
\includegraphics[width=\textwidth]{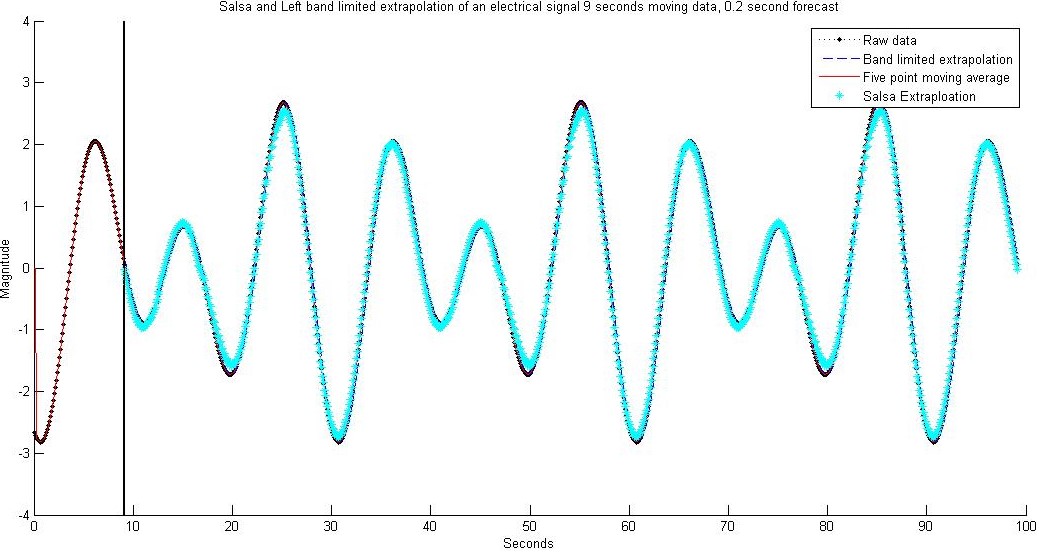}
\caption{Salsa and Causal extrapolation forecasts for 0.2 seconds of an electrical signal repeated for 100 seconds}
\end{figure}
\break

\begin{figure}[h!]
\centering
\includegraphics[width=\textwidth]{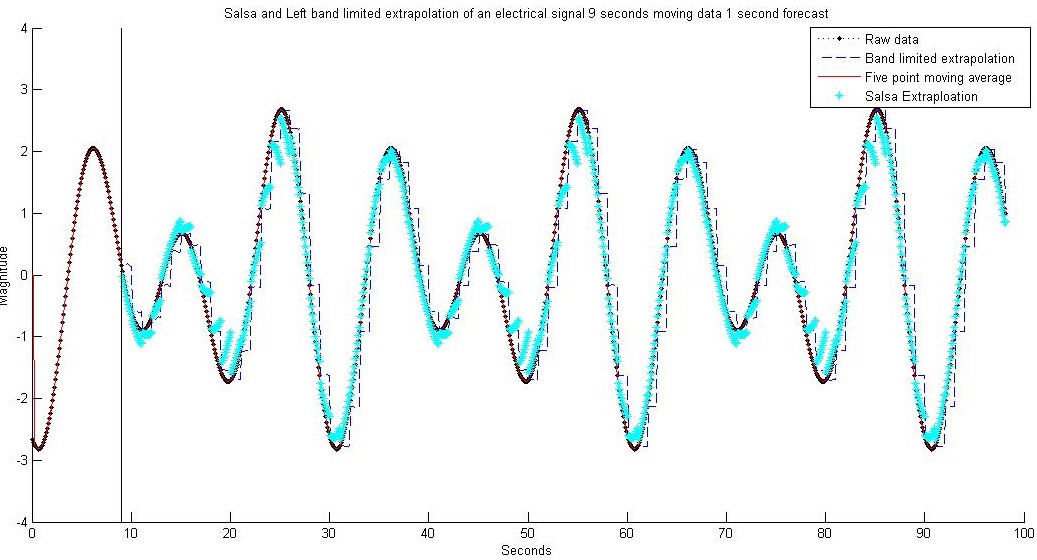}
\caption{Salsa and Causal extrapolation forecasts for 1 seconds of an electrical signal repeated for 100 seconds }
\end{figure}
\clearpage

\section{Bureau of Meteorology Experiements}
Four experiments have been conducted using information from the Australian Bureau of Meteorology, maximum temperature forecasts of the next week made with 91 points of data repeated 38 times during the year, forecasts for the next 2 day with the same data repeated 133 times during the year and both tests repeated for minimum temperature forecasts.
Data from the Australian Bureau of meteorology was retrieved from the “Perth Metro” station located -31.9192 Latitude ,115.8728 Longitude at an altitude of 25.9 meters [4] [5]. Data ranged from the Stations first open in January first 1994 to the twenty fifth of October 2018. This data was selected as it had an estimated 100 completeness for Maximum air temperature data and over 2 decades of collected data as well as being the local weather station. For the purposes of testing the complete year 2017 was used.

\begin{figure}[h!]
\centering
\includegraphics[width=\textwidth]{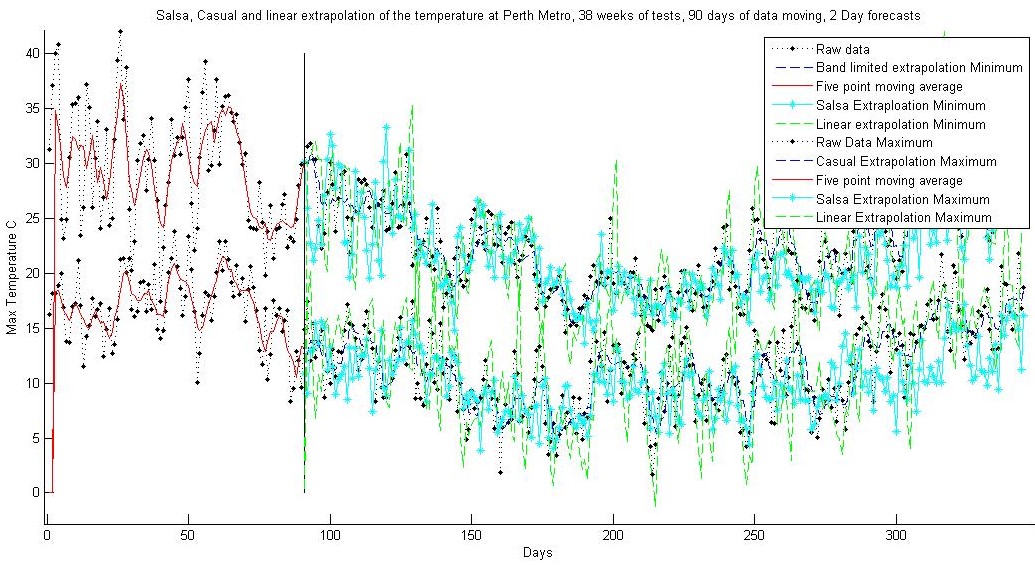}
\caption{Salsa and Causal and linear extrapolation forecasts for 2 days temperature data over the period of a year}
\end{figure}

\begin{table}[h!]
\begin{tabular}{|c|c|c|c||c|c|c|c|}
 \hline
 \multirow{2}{2em}{Data Type}& \multirow{2}{2em}{Raw Data}&\multirow{2}{6em}{Casual Extrapolation}&\multirow{2}{6em}{Salsa Extrapolation}&\multirow{2}{5em}{Comparison method}&\multirow{2}{5em}{Casual Forecast}&\multirow{2}{5em}{Salsa Forecast}&\multirow{2}{5em}{Linear Forecast} \\
&&&&&&&\\
 \hline
Min &14.2 &15.6&13.66&\multirow{2}{4em}{Total L2 residual}&\multirow{2}{4em}{2.1800e03}&\multirow{2}{4em}{4.8381e03}&\multirow{2}{4em}{712105e03} \\
\cline{1-4}
 Max&37.7&33.6&33.32& & & &\\
\hline
Mean &23.05&22.95&21.66&\multirow{3}{4em}{Total L2 residual per point}&\multirow{3}{4em}{8.5492}&\multirow{3}{4em}{18.9730}&\multirow{3}{4em}{27.9256} \\
\cline{1-4}
STD &5.054&4.320&3.942&&&&\\
\cline{1-4}
Range &23.5&18&19.65&&&&\\
\hline
\end{tabular}
\caption{Statistical Results of BOM 2 Day Maximum temperature forecasts}
\end{table}

\begin{table}[h!]
\begin{tabular}{|c|c|c|c||c|c|c|c|}
 \hline
 \multirow{2}{2em}{Data Type}& \multirow{2}{2em}{Raw Data}&\multirow{2}{6em}{Casual Extrapolation}&\multirow{2}{6em}{Salsa Extrapolation}&\multirow{2}{5em}{Comparison method}&\multirow{2}{5em}{Casual Forecast}&\multirow{2}{5em}{Salsa Forecast}&\multirow{2}{5em}{Linear Forecast} \\
&&&&&&&\\
 \hline
Min &1.7&5.38&3.915&\multirow{2}{4em}{Total L2 residual}&\multirow{2}{4em}{1.8570e03}&\multirow{2}{4em}{3.7586e03}&\multirow{2}{4em}{6.1947e03} \\
\cline{1-4}
 Max&21.8&18.31&17.21& & & &\\
\hline
Mean &11.43&11.37&10.18&\multirow{3}{4em}{Total L2 residual per point}&\multirow{3}{4em}{7.2822}&\multirow{3}{4em}{14.7395}&\multirow{3}{4em}{24.2931} \\
\cline{1-4}
STD &3.902&3.12&2.766&&&&\\
\cline{1-4}
Range &20.1&12.93&13.3&&&&\\
\hline
\end{tabular}
\caption{Statistical Results of BOM 2 Day Minimum temperature forecasts}
\end{table}

\begin{table}[h!]
\begin{tabular}{|c|c|c|c||c|c|c|c|}
 \hline
 \multirow{2}{2em}{Data Type}& \multirow{2}{2em}{Raw Data}&\multirow{2}{6em}{Casual Extrapolation}&\multirow{2}{6em}{Salsa Extrapolation}&\multirow{2}{5em}{Comparison method}&\multirow{2}{5em}{Casual Forecast}&\multirow{2}{5em}{Salsa Forecast}&\multirow{2}{5em}{Linear Forecast} \\
&&&&&&&\\
 \hline
Min &14.2&16.11&13.53&\multirow{2}{4em}{Total L2 residual}&\multirow{2}{4em}{3.3970e03}&\multirow{2}{4em}{6.5986e03}&\multirow{2}{4em}{8.13195e03} \\
\cline{1-4}
 Max&37.7&32.83&33.87& & & &\\
\hline
Mean &23.25&23.31&21.67&\multirow{3}{4em}{Total L2 residual per point}&\multirow{3}{4em}{12.7229}&\multirow{3}{4em}{24.7137}&\multirow{3}{4em}{30.4564} \\
\cline{1-4}
STD &5.104&4.482&4.352&&&&\\
\cline{1-4}
Range &23.5&16.71&20.35&&&&\\
\hline
\end{tabular}
\caption{Statistical Results of BOM 7 Day Maximum temperature forecasts}
\end{table}

\begin{table}[h!]
\begin{tabular}{|c|c|c|c||c|c|c|c|}
 \hline
 \multirow{2}{2em}{Data Type}& \multirow{2}{2em}{Raw Data}&\multirow{2}{6em}{Casual Extrapolation}&\multirow{2}{6em}{Salsa Extrapolation}&\multirow{2}{5em}{Comparison method}&\multirow{2}{5em}{Casual Forecast}&\multirow{2}{5em}{Salsa Forecast}&\multirow{2}{5em}{Linear Forecast} \\
&&&&&&&\\
 \hline
Min &1.7&5.685&3.725&\multirow{2}{4em}{Total L2 residual}&\multirow{2}{4em}{2.7701e03}&\multirow{2}{4em}{5.0055e03}&\multirow{2}{4em}{6.3337e03} \\
\cline{1-4}
 Max&21.8&17.6&17.28& & & &\\
\hline
Mean &11.63&11.56&9.845&\multirow{3}{4em}{Total L2 residual per point}&\multirow{3}{4em}{10.3750}&\multirow{3}{4em}{18.7474}&\multirow{3}{4em}{23.7216} \\
\cline{1-4}
STD &3.937&3.14&2.504&&&&\\
\cline{1-4}
Range &20.1&11.91&13.56&&&&\\
\hline
\end{tabular}
\caption{Statistical Results of BOM 7 Day Minimum temperature forecasts}
\end{table}

\begin{figure}[h!]
\centering
\includegraphics[width=\textwidth]{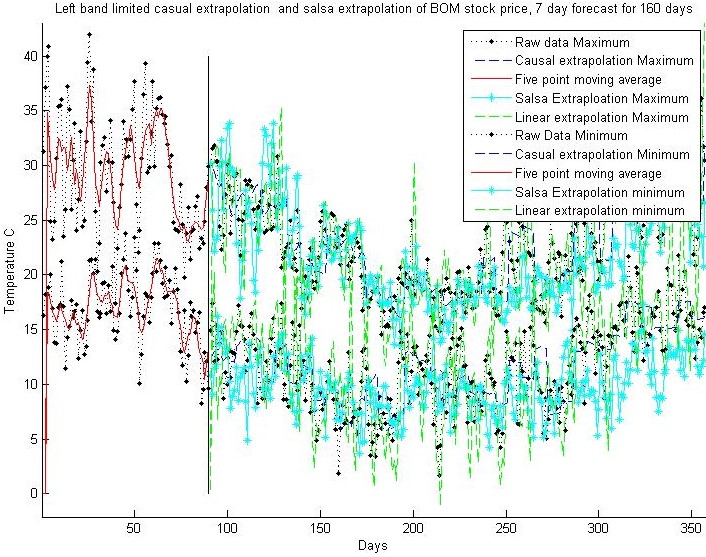}
\caption{Salsa and Causal and linear extrapolation forecasts for 7 days temperature data over the period of a year}
\end{figure}

In these experiments the superiority of Casual smoothing extrapolation is clear with the L2- euclidean in the temperature data being 45-55\% smaller then its nearest competitor.Though while its predictions are far closer to the actual points then the other two methods it compresses the data range to 65-77\% of the raw data range. Comparitivly the Salsa extrapolation compresses the data to 67-84\% of the raw range. While both perform significantly better linear extrapolation and neither have a forecast so poor that the maximum temperature forecast is below the minimum temperature for the day of vice versa for both experiements. Casual smoothing extrapolation appears to be a significantly more useful forecasting tool for this particular data set.
\break
\section{Australian stock exchange experiements}
Similar to the previous section 4 tests were conducted using data from the Australian stock exchange, with 2 day and 5-day forecasts of minimum and maximum prices repeated 79 and 31 respectfully with moving 91-point data sets. The data set for the testing of stock prices were obtained from Market index [6]consisting of the opening, high, low and close values of a tradeable stock and index fund between the 11th of October 2017 and the 10th of October 2018. These values were also confirmed with Commonwealth Securities Limited quotes [7] of these data points.

\begin{figure}[h!]
\centering
\includegraphics[width=\textwidth]{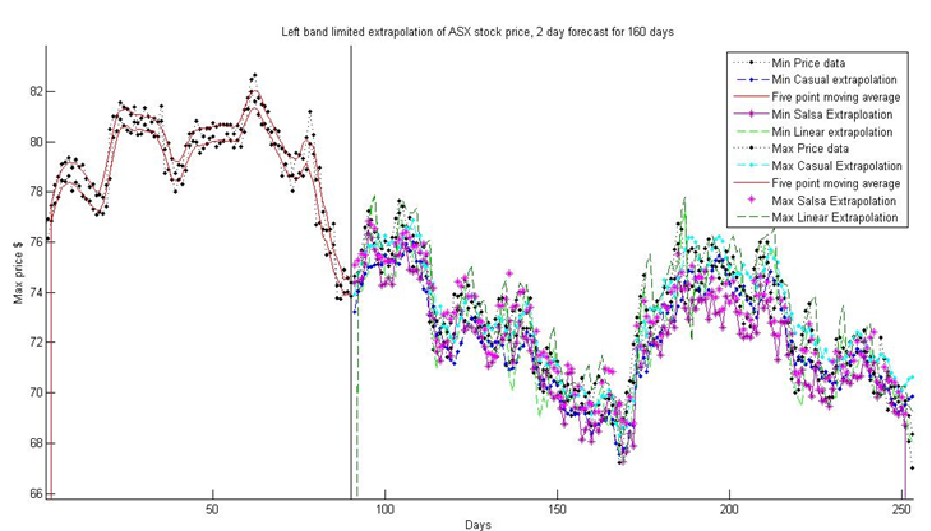}
\caption{Salsa and Causal and linear extrapolation forecasts for 2 days ASX stock data over the period of a year}
\end{figure} 

\begin{table}[h!]
\begin{tabular}{|c|c|c|c||c|c|c|c|}
 \hline
 \multirow{2}{2em}{Data Type}& \multirow{2}{2em}{Raw Data}&\multirow{2}{6em}{Casual Extrapolation}&\multirow{2}{6em}{Salsa Extrapolation}&\multirow{2}{5em}{Comparison method}&\multirow{2}{5em}{Casual Forecast}&\multirow{2}{5em}{Salsa Forecast}&\multirow{2}{5em}{Linear Forecast} \\
&&&&&&&\\
 \hline
Min &67.92&68.34&68.8&\multirow{2}{4em}{Total L2 residual}&\multirow{2}{4em}{150.9761}&\multirow{2}{4em}{194.0201}&\multirow{2}{4em}{5.8375e03} \\
\cline{1-4}
 Max&77.66&76.86&76.82 & & &\\
\hline
Mean &73.12&72.98&71.86&\multirow{3}{4em}{Total L2 residual per point}&\multirow{3}{4em}{0.9320}&\multirow{3}{4em}{1.2437}&\multirow{3}{4em}{36.034} \\
\cline{1-4}
STD &70.25&2.13&72.88&&&&\\
\cline{1-4}
Range &2.242&8.52&8.02&&&&\\
\hline
\end{tabular}
\caption{Statistical Results of ASX 2 Day Maximum price forecasts}
\end{table}

\begin{table}[h!]
\begin{tabular}{|c|c|c|c||c|c|c|c|}
 \hline
 \multirow{2}{2em}{Data Type}& \multirow{2}{2em}{Raw Data}&\multirow{2}{6em}{Casual Extrapolation}&\multirow{2}{6em}{Salsa Extrapolation}&\multirow{2}{5em}{Comparison method}&\multirow{2}{5em}{Casual Forecast}&\multirow{2}{5em}{Salsa Forecast}&\multirow{2}{5em}{Linear Forecast} \\
&&&&&&&\\
 \hline
Min &-67&67.56&67.24&\multirow{2}{4em}{Total L2 residual}&\multirow{2}{4em}{150.3173}&\multirow{2}{4em}{195.6079}&\multirow{2}{4em}{5.7525e03} \\
\cline{1-4}
 Max&77.17&75.99&76.32& & & &\\
\hline
Mean &72.18&72.05&70.97&\multirow{3}{4em}{Total L2 residual per point}&\multirow{3}{4em}{0.9279}&\multirow{3}{4em}{1.2539}&\multirow{3}{4em}{35.5091} \\
\cline{1-4}
STD &2.237&2.119&8.179&&&&\\
\cline{1-4}
Range &10.17&8.437&9.08&&&&\\
\hline
\end{tabular}
\caption{Statistical Results of ASX 2 Day Minimum price forecasts}
\end{table}

\begin{table}[h!]
\begin{tabular}{|c|c|c|c||c|c|c|c|}
 \hline
 \multirow{2}{2em}{Data Type}& \multirow{2}{2em}{Raw Data}&\multirow{2}{6em}{Casual Extrapolation}&\multirow{2}{6em}{Salsa Extrapolation}&\multirow{2}{5em}{Comparison method}&\multirow{2}{5em}{Casual Forecast}&\multirow{2}{5em}{Salsa Forecast}&\multirow{2}{5em}{Linear Forecast} \\
&&&&&&&\\
 \hline
Min &67.92&68.83&68.8&\multirow{2}{4em}{Total L2 residual}&\multirow{2}{4em}{296.9839}&\multirow{2}{4em}{354.9674}&\multirow{2}{4em}{5.8375e03} \\
\cline{1-4}
 Max&77.66&77.64&76.74& & & &\\
\hline
Mean &73.12&73.2&71.66&\multirow{3}{4em}{Total L2 residual per point}&\multirow{3}{4em}{1.8332}&\multirow{3}{4em}{2.1912}&\multirow{3}{4em}{36.034} \\
\cline{1-4}
STD &2.242&2.154&8.25&&&&\\
\cline{1-4}
Range &9.74&8.807&7.94&&&&\\
\hline
\end{tabular}
\caption{Statistical Results of ASX 5 Day Maximum price forecasts}
\end{table}

\begin{table}[h!]
\begin{tabular}{|c|c|c|c||c|c|c|c|}
 \hline
 \multirow{2}{2em}{Data Type}& \multirow{2}{2em}{Raw Data}&\multirow{2}{6em}{Casual Extrapolation}&\multirow{2}{6em}{Salsa Extrapolation}&\multirow{2}{5em}{Comparison method}&\multirow{2}{5em}{Casual Forecast}&\multirow{2}{5em}{Salsa Forecast}&\multirow{2}{5em}{Linear Forecast} \\
&&&&&&&\\
 \hline
Min &67&68.02&67.9&\multirow{2}{4em}{Total L2 residual}&\multirow{2}{4em}{283.3359}&\multirow{2}{4em}{315.1025}&\multirow{2}{4em}{5.7525e03} \\
\cline{1-4}
 Max&77.17&76.89&76.03& & & &\\
\hline
Mean &72.18&72.26&70.89&\multirow{3}{4em}{Total L2 residual per point}&\multirow{3}{4em}{1.749}&\multirow{3}{4em}{1.9451}&\multirow{3}{4em}{35.5091} \\
\cline{1-4}
STD &2.237&2.167&8.163&&&&\\
\cline{1-4}
Range &10.17&8.871&8.13&&&&\\
\hline
\end{tabular}
\caption{Statistical Results of ASX 5 Day Minimum price forecasts}
\end{table}

\begin{figure}[h!]
\centering
\includegraphics[width=\textwidth]{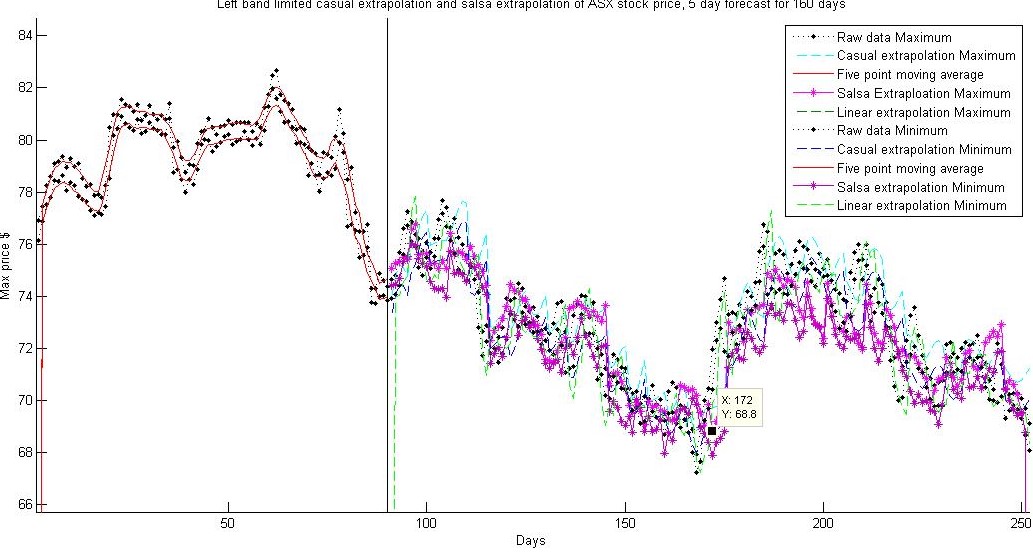}
\caption{Salsa and Causal and linear extrapolation forecasts for 5 days ASX stock data over the period of a year}
\end{figure}

This series of experiments had similar results to the temperature forecasts with Causal extrapolation being the superior forecasters with the most compressed data set. The main difference being that in three out of the four cases the Causal extrapolation had ranges closer to raw data range than the SALSA extrapolation, forecasting data which visually resembles the shape of the underlying process more. The SALSA extrapolated points also had forecasted maximum prices which fell below the minimum price forecasted and minimum forecasted prices which were above the maximum price forecasted, while Casual smoothing extrapolation did not. There for if SALSA extrapolation was used to forecast prices in an attempt to purchase and sell stocks during a 2- or 5-day periods you would have a maximum price falling lower than its minimum price resulting in a catastrophic loss if trades were made with an autonomous system. Causal smoothing extrapolation does not have as large a chance of experiencing this catastrophic failure because at no point does its forecast for the maximum price fall lower than the forecast for the minimum price or vice versa. For this data type the more conservative and reliable forecasts made by casual smoothing extrapolation is clearly the better choice.

\section{Summary and Conclusion}
In all cases the primary methods of forecasting discussed in this paper are superior to a basic linear forecast when the underlying processes are unknown and Salsa extrapolation is clearly superior in the known process. In the case of the complex underlying process Casual Smoothing extrapolation had a lower L2 euchlideon then Salsa extrapolation, resulting in a forecast more likely to be closer to the actual value that will occur. This is ideal for the Australian Stock exchange data as it will result in the minimum amount of loss when conducting trades and making trades closer to the minimum and maximum prices on any given day. It also has not produced a forecast in which a loss would occur if a stock was sold at the maximum forecast price and bought at the minimum forecast price, while this has occurred multiple times with SALSA extrapolation. This is not to say that either processes would be guaranteed to make a profit on any give day or over time, but the Salsa extrapolation has the greatest capacity to incur a long-term loss or catastrophic short-term loss for an autonomous trading process.
In the case of the weather data which is the superior method of prediction is not as straight forward as no cross over of the data between minimum and maximum forecasts occurs. Though the casual smoothing extrapolation had a L2 euclidean half of the SALSA forecasts and a range at most 82\% of the SALSA range. This would result in the Casual smoothing forecasts having far more accurate forecasts of the expected maximum and minimum temperatures for any given day while the SALSA forecast would produce a winder band of temperatures which could prove more useful for preparing for what kind of weather tomorrow may bring.
The forecasting of electrical signals is the most simple series to discuss as the SALSA forecasts are more accurate, barely compressed and are completed faster than Casual smoothing extrapolation. This makes Casual smoothing extrapolation a redundent method for forecasting electrical signals as forecast period was shorter then the time it took to complete a prediction. SALSA extrapolation on the other hand is shown to be a forecasting processes that is highly practical for making short range predicitions of electrical signal data. This lead to the conclusion that Casual smoothing extrapolation is a method useful when the underlying process does not have forecasting method or is too complex to reasonably forecast but does not out perform a specially tailored forecast solution. 
These experiements that have been conducted cannot be considered exhaustive, and it would be advisable to do more extensive tests on each data set, adjust the parameters for both SALSA and Causal smoothing extrapolation depending on the data set and adjusting it depending on fluctuation in that set. This would achieve more accurate results which are unattainable in the confines of this paper. The most conclusive statement that can be made about these forecasting methods is that Causal smoothing extrapolation provides a more conservative estimation of future series data while losing more information about said data set. While SALSA extrapolations are in general less likely to be the actual values that occur in the future, but are much faster and have a wider range of values. With the exeption of the process which it has been designed to forecast, in which it is the superior forecasting method in all aspects. Each of these characteristics could be more useful depending on circumstances and the need for forecasting. Finally both are far better than following an average trend line when the underlying process is either too complex or unknown.
\section*{References}\
\hphantom{xx} [1]  N. Dokuchaev, "On causal extrapolation of sequences with applications to forecasting,
Applied Mathematics and Computation, vol. 328, pp. 276-286, 2018. 
\par [2] 	N. Rowe, "Applications of band-limited extrapolation to forecasting of weather 
and financial time series," arXiv, N/A, 2019.
\par [3] 	M. V. Afonso, J. M. Bioucas-Dias and M. A. T. Figueriredo, "Fast Image Recovery Using
 Variable Splitting and Constrained Optimization," IEEE Transcation on Image Processing,
 vol. 19, no. 9, pp. 2345-2356, 2010. 
\par [4] 	Australian Government Bureau of Meterorology , "Australian Government Bureau of 
Meterorology,"  http://www.bom.gov.au/jsp/ncc/cdio/weatherData
/av?p\_nccObsCode=122\&p\_display\_type=dailyDataFile\&p\_startYear=\&p\_c=\&p\_stn\_num=009225. 
 [Accessed 11 10 2018].
\par [5] 	Australian Government Bureau of Meterology, "Basic Climatological Station Metadata," 
26 7 2018. 
 http://www.bom.gov.au/clim\_data/cdio/metadata/pdf/siteinfo/IDCJMD0040.009225.SiteInfo.pdf. 
\par [Accessed 31 10 2018].
\par [6] 	Market Index., "S\&P/ASX 200," 10 10 2018. 
https://www.marketindex.com.au/asx200. [Accessed 10 10 2018].

\par [7] 	Commonwealth Securities Limited, "Trade History CBA," 26 10 2018. 
https://www2.commsec.com.au/Private/MarketPrices/TradeHistory
/TradeHistory.aspx?stockCode=CBA. [Accessed 2018 10 26].
\par [8] 	Australian Government Department of Enviorment and energy, "Portfolio Budget Statements
 2017-18," http://www.environment.gov.au/about-us/publications/budget/portfolio-budget-
statements-2017-18. [Accessed 2 11 2018].
\par [9] 	N. Bingham, "Szegö’s theorem and its probabilistic descendants," Probab. Surveys 9 ,
 vol. 9, pp. 287-324, 2012. 
\par [10] 	N. Dokuchaev, "Filtration and extrapolation in trajectory analysis based on 
spectrum methods," Working paper, 2019.
\par [11] 	I. Selesnick, "Introduction to Sparsity in Signal Processing," Ivan Selesnick, 2012.
\par [12] 	I. Selesnick, "L1-Norm Penalized Least Squares with Salsa," Connexions, 2014.
\newpage
\setcounter{equation}{0}
\renewcommand{\thesubsection}{A.\arabic{subsection}}

\renewcommand{\theequation}{A.\arabic{equation}}
\renewcommand{\thelemma}{A.\arabic{lemma}}
\renewcommand{\theproposition}{A.\arabic{proposition}}

\section*{Appendix}
\subsection{Monte Carlo Extrapolation}
\begin{verbatim}
%Start by creating Monte-Carlo random numbers 1-D
%Z(t)=A(t)Z(t-1)+o*u(t)
%Input carlosim(N,o,z0,v) for N for 2N+1 points, z0 initial points, v
%dimensions, o>0 constant.
%Out puts A = v*v*N, u=v*1*N, z=v*1*N where z is the x(t) values
clear all
muset=zeros(2,200);
mu=0.6
NIT=1000;
lambda= 1;
tots=0;
cost=zeros(1,NIT);
time=0
for j=0:1:1000
[B,u,z]=carlosim(55,1,1,1);
S=0;
M=98;
N=200;
y=z(1+S:M+S)'+80;
% Define transform
% Oversampled DFT  
% N : FFT length (including zero-=padding)
%Define Nit 
%K =91 Length of observed signal
%Creates vector s of length M with K values in order given by vector k
K=91;
k=ones(M,1);
k=k(1:K);
s=true(1,K)';
s(1:K)=true;
s(K+1:M)=false;
Y=y;
Y(~s)=0;
p=200;
c =AT(Y,M,N);
x1=AT(Y,M,N);
size(c);
size(A(c,M,N));
d = zeros(size(c));
cost = zeros(1,NIT);
for i = 1:NIT
    u = soft(c + d, lambda/mu) - d;
    d = 1/(mu+p) * AT((Y - s.*A(u,M,N)),M,N);
    c = d + u;
    residual = Y - A(c,M,N);
    cost(i) = sum(abs(residual(:)).^2) + sum(abs(lambda * c(:)));
end
x=A(c,M,N);
tots=tots+sum(abs(((z(91:98)'+80)-x(91:98))).^2);
x=0;
time=time+1
end
tots
 
cost(NIT)
t=1:98;
x=A(c,M,N)
hold on
plot(t(1:91),real(x(1:91)),'gx')
plot(t(91:98),real(x(91:98)),'rx')
plot(t(1:98),z(1:98)+80,'b')
title('Simulation of Salsa Extrapolation')
xlabel('N')
ylabel('Magnitude')
line([0,0],[-3,3],'Color',[0,0,0],'HandleVisibility','off')
hold all
legend('Salsa data points','Extrapolated Points','Original simulation points')
\end{verbatim}
\subsection{Electronic signal forecast with Salsa and Causal Extrapolation}
\begin{verbatim}
%Start by importing raw data
AA=load('C:\Users\valough\Desktop\MATH5001\P.mat');
z=AA.SIG(1,:);
N=45;
n=-N:1:N;
M=2*N+1;
X=pi;
%Currently unused
s=91;
q=1;
ts=q:1:s;
time=0
%Moving average
MV=zeros(length(z),1);
MV(1)=(z(1)+z(2)+z(3)+z(4)+z(5))/5;
MV(1)=MV(2);
MV(1)=MV(3);
MV(length(z)-1)=(z(length(z)-1)+z(length(z)-2)+z(length(z)-3)+z(length(z)-4)+z(length(z)))/5;
MV(length(z)-1)=MV(length(z)-2);
MV(length(z)-1)=MV(length(z)-3);
for l=3:(length(z)-3)
    MV(l)=(z(l)+z(l+1)+z(l+2)+z(l-2)+z(l-1))/5;
end
MA=mean(MV(5:length(MV)-5));
LE=zeros(10,1);
for tt=91:10:991
    mm=(z(tt)-z(tt-1))/10;
    cc=(z(tt)-((z(tt)-z(tt-1))/10)*tt);
    for it=1:10
        LE(tt+it-90)=mm*(tt+it)+cc;
    end    
end
%% SALSA FORECAST
smu=0.6;
stots=0;
sM=101;
sN=200;
sNIT=1000;
slambda=1;
sp=200;
sS=0;
sD=zeros(163,1);
timesalsa=0
for sS=0:10:890
    sy=z(1+sS:sM+sS)';
    sK=91;
    sk=ones(sM,1);
    sk=sk(1:sK);
    ss=true(1,sK)';
    ss(1:sK)=true;
    ss(sK+1:sM)=false;
    sY=sy;
    sY(~ss)=0;
    sc =AT(sY,sM,sN);
    sx1=AT(sY,sM,sN);
    sd = zeros(size(sc));
    cost = zeros(1,sNIT);
     for i = 1:sNIT
        su = soft(sc + sd, 0.5*slambda/smu) - sd;
        sd = 1/(smu+sp) * AT((sY - ss.*A(su,sM,sN)),sM,sN);
        sc = sd + su;
        residual = sY - A(sc,sM,sN);
        cost(i) = sum(abs(residual(:)).^2) + sum(abs(slambda * sc(:)));
    end
    sx=A(sc,sM,sN);
    stots=stots+sum(abs(((z(92+sS:101+sS)')-sx(92:101))).^2);
    sD(sS+1)=sx(92);
    sD(sS+2)=sx(93);
    sD(sS+3)=sx(94);
    sD(sS+4)=sx(95);
    sD(sS+5)=sx(96);
    sD(sS+6)=sx(97);
    sD(sS+7)=sx(98);
    sD(sS+8)=sx(99);
    sD(sS+9)=sx(100);
    sD(sS+10)=sx(101);
    timesalsa=timesalsa+1
    
end
sD
size(sD)
%% LEFT BAND LIMITED CASUAL EXTRAPOLATION
length(LE(1:163))
length(91:1:253)
%Process Q*R*Q+*x(t)
%Assign omega=g 
g=pi/4;
%Creates list of Q+*x(t)
V=zeros(25,1);
for ii=0:89;
    QX=zeros(M,1);
    MEAN=mean(MV((89+10*ii):(91+10*ii)));
    MEAN2=mean(MV((1+10*ii):(91+10*ii)));
    MV((1+10*ii):(91+10*ii))=MV((1+10*ii):(91+10*ii))-MEAN2;
    for k=-N:1:N
        QX(k+N+1)=0;
        for t=ts;
            QX(k+N+1)=QX(k+N+1)+(g/pi)*sinc((k*pi+g*t)/X)*MV(t);
        end
    end
    %Creates matrix Rkm
    R=zeros(M,M);
    for k=-N:1:N
        for m=-N:1:N
            for t=ts
            R(k+N+1,m+N+1)=R(k+N+1,m+N+1)+(g/pi)^2*sinc((k*pi+g*t)/X)*sinc((m*pi+g*t)/X);
            end
        end
    end
    %where v is epsilon
    v=0.1;
    RI= R+eye(M)*v;
    y=inv(RI)*QX;
    %Creates list of Q sum of all Yk for each t
    Q=zeros(length(ts)+14,1);
    for t=q:1:(s+12)
        for k=-N:1:N
            Q(t-ii*10)=Q(t-ii*10)+y(k+N+1)*(g/pi)*sinc((k*pi+g*t)/X);
        end
    end
    %Calulating residual
    t=q:1:s+12;
    BLR=0;
    D=zeros(length(Q),1);       
    D=Q(:);
    %for i=1:1:length(Q)
         %for j=1:1:12
             %if zm(i)==j
              %  D(i)=D(i)+S(j);
             %       end
 
        % end
    %end
    for i=1:1:91
        BLR=BLR+sqrt((z(i+10*ii)-(D(i)+MEAN2))^2);
    end
    MV((1+10*ii):(91+10*ii))=MV((1+10*ii):(91+10*ii))+MEAN2;
    BLR
    BLR/90
    D(1:91)=D(1:91)+MEAN2;
    MAA=mean(D(1:91));
    D(92:104)=D(92:104)+MEAN;
    V(q:q+12)=D(92:104);
    s=s+10;
    q=q+10;
    time=time+1
    BLR=0;
end
BLRE=0;
 
%% Residual calculation
for i=1:1:891
    BLRE=BLRE+abs((z(i+91)-(V(i)))^2);
end
LERR=0;
for i=1:1:891
    LERR=LERR+abs((z(i+91)-(LE(i)))^2);
end
length(V);
BLRE
BLRE/length(V(1:891))
LERR
LERR/length(V(1:891))
stots
stots/length(sD(1:891))
length(LE);
length(V);
length(z);
length(MV);
length(sD);
tt=9.1:0.1:98.1;
length(tt);
 
hold on
plot(0.1:0.1:9.1,z(1:91),'k.:',0.1:0.1:9.1,MV(1:91),'r','HandleVisibility','off')
plot(tt,z(91:981),'k.:',tt,V(1:891),'b--',tt,MV(91:981),'r',tt,sD(1:891),'c*')
title('Salsa and Left band limited extrapolation of an electrical signal 9 seconds
 moving data')
xlabel('Seconds')
ylabel('Magnitude')
line([9,9],[-4,4],'Color',[0,0,0],'HandleVisibility','off')
hold all
plot(tt,z(91:981),'k.:',tt,V(1:891),'b--',tt,MV(91:981),'r',tt,sD(1:891),'c-*')
legend('Raw data','Band limited extrapolation','Five point moving average','Salsa 
Extraploation')
\end{verbatim}
\subsection{BOM forecast with Salsa, Causal and Linear Extrapolation}
\begin{verbatim}
%Start by importing raw data
AA=xlsread('C:\Users\valough\Documents\MATLAB\IDCJAC0010_009225_2017_Data.csv');
z=transpose(AA(:,5));
zm=transpose(AA(:,3));
N=45;
n=-N:1:N;
M=2*N+1;
X=pi;
%Currently unused
s=91;
q=1;
ts=q:1:s;
time=0
%Moving average
MV=zeros(length(z),1);
MV(1)=(z(1)+z(2)+z(3)+z(4)+z(5))/5;
MV(1)=MV(2);
MV(1)=MV(3);
MV(length(z)-1)=(z(length(z)-1)+z(length(z)-2)+z(length(z)-3)+z(length(z)-4)+z(length(z)))/5;
MV(length(z)-1)=MV(length(z)-2);
MV(length(z)-1)=MV(length(z)-3);
for l=3:(length(z)-3)
    MV(l)=(z(l)+z(l+1)+z(l+2)+z(l-2)+z(l-1))/5;
end
MA=mean(MV(5:length(MV)-5));
LE=zeros(10,1);
for tt=91:2:343
    mm=(z(tt)-z(tt-1))/2;
    cc=(z(tt)-((z(tt)-z(tt-1))/2)*tt);
    for it=1:2
        LE(tt+it-90)=mm*(tt+it)+cc;
    end    
end
%% SALSA FORECAST
smu=0.6;
stots=0;
sM=94;
sN=200;
sNIT=1000;
slambda=1;
sp=200;
sS=0;
sD=zeros(163,1);
timesalsa=0
for sS=0:2:266
    sy=z(1+sS:sM+sS)';
    sK=91;
    sk=ones(sM,1);
    sk=sk(1:sK);
    ss=true(1,sK)';
    ss(1:sK)=true;
    ss(sK+1:sM)=false;
    sY=sy;
    sY(~ss)=0;
    sc =AT(sY,sM,sN);
    sx1=AT(sY,sM,sN);
    sd = zeros(size(sc));
    cost = zeros(1,sNIT);
     for i = 1:sNIT
        su = soft(sc + sd, 0.5*slambda/smu) - sd;
        sd = 1/(smu+sp) * AT((sY - ss.*A(su,sM,sN)),sM,sN);
        sc = sd + su;
        residual = sY - A(sc,sM,sN);
        cost(i) = sum(abs(residual(:)).^2) + sum(abs(slambda * sc(:)));
    end
    sx=A(sc,sM,sN);
    stots=stots+sum(abs(((z(92+sS:93+sS)')-sx(92:93))).^2);
    sD(sS+1)=sx(92);
    sD(sS+2)=sx(93);
    timesalsa=timesalsa+1
    
end
sD(sS+3)=sx(94);
sD
size(sD)
%% LEFT BAND LIMITED CASUAL EXTRAPOLATION
length(LE(1:255))
length(91:1:345)
%Process Q*R*Q+*x(t)
%Assign omega=g 
g=pi/4;
%Creates list of Q+*x(t)
V=zeros(25,1);
for ii=0:133;
    QX=zeros(M,1);
    MEAN=mean(MV((89+2*ii):(91+2*ii)));
    MEAN2=mean(MV((1+2*ii):(91+2*ii)));
    MV((1+2*ii):(91+2*ii))=MV((1+2*ii):(91+2*ii))-MEAN2;
    for k=-N:1:N
        QX(k+N+1)=0;
        for t=ts;
            QX(k+N+1)=QX(k+N+1)+(g/pi)*sinc((k*pi+g*t)/X)*MV(t);
        end
    end
    %Creates matrix Rkm
    R=zeros(M,M);
    for k=-N:1:N
        for m=-N:1:N
            for t=ts
            R(k+N+1,m+N+1)=R(k+N+1,m+N+1)+(g/pi)^2*sinc((k*pi+g*t)/X)*sinc((m*pi+g*t)/X);
            end
        end
    end
    %where v is epsilon
    v=0.1;
    RI= R+eye(M)*v;
    y=inv(RI)*QX;
    %Creates list of Q sum of all Yk for each t
    Q=zeros(length(ts)+14,1);
    for t=q:1:(s+6)
        for k=-N:1:N
            Q(t-ii*2)=Q(t-ii*2)+y(k+N+1)*(g/pi)*sinc((k*pi+g*t)/X);
        end
    end
    %Calulating residual
    t=q:1:s+6;
    BLR=0;
    D=zeros(length(Q),1);       
    D=Q(:);
    %for i=1:1:length(Q)
         %for j=1:1:12
             %if zm(i)==j
              %  D(i)=D(i)+S(j);
             %       end
 
        % end
    %end
    for i=1:1:91
        BLR=BLR+sqrt((z(i+2*ii)-(D(i)+MEAN2))^2);
    end
    MV((1+2*ii):(91+2*ii))=MV((1+2*ii):(91+2*ii))+MEAN2;
    BLR
    BLR/90
    D(1:91)=D(1:91)+MEAN2;
    MAA=mean(D(1:91));
    D(92:98)=D(92:98)+MEAN;
    V(q:q+6)=D(92:98);
    s=s+2;
    q=q+2;
    time=time+1
    BLR=0;
end
BLRE=0;
 
%% Residual calculation
for i=1:1:255
    BLRE=BLRE+abs((z(i+91)-(V(i)))^2);
end
LERR=0;
for i=1:1:255
    LERR=LERR+abs((z(i+91)-(LE(i)))^2);
end
length(V);
BLRE
BLRE/length(V(1:255))
LERR
LERR/length(V(1:255))
stots
stots/length(sD(1:255))
length(LE(1:255));
length(V(1:255));
length(z(91:345));
length(MV(91:345));
tt=91:1:345;
length(tt);
 
hold on
plot(1:91,z(1:91),'k.:',1:91,MV(1:91),'r','HandleVisibility','off')
plot(tt,z(91:345),'k.:',tt,V(1:255),'b--',tt,MV(91:345),'r',tt,sD(1:255),'c*')
title('Salsa and Left band limited extrapolation of the temperature at Perth Metro, 38 weeks 
of tests, 90 days of data moving')
xlabel('Days')
ylabel('Max Temperature C')
line([91,91],[0,40],'Color',[0,0,0],'HandleVisibility','off')
hold all
plot(91:1:345,LE(1:255),'g--')
legend('Raw data','Band limited extrapolation','Five point moving average','Salsa
 Extraploation','Linear extrapolation')
\end{verbatim}
\subsection{ASX forecast with Salsa, Causal and Linear Extrapolation}
\begin{verbatim}
%Start by importing raw data
AA=xlsread('C:\Users\valough\Documents\MATLAB\CBA.AX.csv');
z=transpose(AA(:,3));
zm=transpose(AA(:,3));
N=45;
n=-N:1:N;
M=2*N+1;
X=pi;
%Currently unused
s=91;
q=1;
ts=q:1:s;
time=0
%Moving average
MV=zeros(length(z),1);
MV(1)=(z(1)+z(2)+z(3)+z(4)+z(5))/5;
MV(1)=MV(2);
MV(1)=MV(3);
MV(length(z)-1)=(z(length(z)-1)+z(length(z)-2)+z(length(z)-3)+z(length(z)-4)+z(length(z)))/5;
MV(length(z)-1)=MV(length(z)-2);
MV(length(z)-1)=MV(length(z)-3);
for l=3:(length(z)-3)
    MV(l)=(z(l)+z(l+1)+z(l+2)+z(l-2)+z(l-1))/5;
end
MA=mean(MV(5:length(MV)-5));
LE=zeros(10,1);
for tt=91:2:253
    mm=(z(tt)-z(tt-1))/2;
    cc=(z(tt)-((z(tt)-z(tt-1))/2)*tt);
    for it=1:2
        LE(tt+it-90)=mm*(tt+it)+cc;
    end    
end
%% SALSA FORECAST
smu=15;
stots=0;
sM=98;
sN=200;
sNIT=1000;
slambda=1;
sp=200;
sS=0;
sD=zeros(163,1);
timesalsa=0
for sS=0:5:155
    sy=z(1+sS:sM+sS)';
    sK=91;
    sk=ones(sM,1);
    sk=sk(1:sK);
    ss=true(1,sK)';
    ss(1:sK)=true;
    ss(sK+1:sM)=false;
    sY=sy;
    sY(~ss)=0;
    sc =AT(sY,sM,sN);
    sx1=AT(sY,sM,sN);
    sd = zeros(size(sc));
    cost = zeros(1,sNIT);
     for i = 1:sNIT
        su = soft(sc + sd, 0.5*slambda/smu) - sd;
        sd = 1/(smu+sp) * AT((sY - ss.*A(su,sM,sN)),sM,sN);
        sc = sd + su;
        residual = sY - A(sc,sM,sN);
        cost(i) = sum(abs(residual(:)).^2) + sum(abs(slambda * sc(:)));
    end
    sx=A(sc,sM,sN);
    stots=stots+sum(abs(((z(92+sS:96+sS)')-sx(92:96))).^2);
    sD(sS+1)=sx(92);
    sD(sS+2)=sx(93);
    sD(sS+3)=sx(94);
    sD(sS+4)=sx(95);
    sD(sS+5)=sx(96);
    timesalsa=timesalsa+1
    
end
sD(sS+6)=sx(97);
sD
size(sD)
%% LEFT BAND LIMITED CASUAL EXTRAPOLATION
length(LE(1:163))
length(91:1:253)
%Process Q*R*Q+*x(t)
%Assign omega=g 
g=pi/4;
%Creates list of Q+*x(t)
V=zeros(25,1);
for ii=0:31;
    QX=zeros(M,1);
    MEAN=mean(MV((89+5*ii):(91+5*ii)));
    MEAN2=mean(MV((1+5*ii):(91+5*ii)));
    MV((1+5*ii):(91+5*ii))=MV((1+5*ii):(91+5*ii))-MEAN2;
    for k=-N:1:N
        QX(k+N+1)=0;
        for t=ts;
            QX(k+N+1)=QX(k+N+1)+(g/pi)*sinc((k*pi+g*t)/X)*MV(t);
        end
    end
    %Creates matrix Rkm
    R=zeros(M,M);
    for k=-N:1:N
        for m=-N:1:N
            for t=ts
            R(k+N+1,m+N+1)=R(k+N+1,m+N+1)+(g/pi)^2*sinc((k*pi+g*t)/X)*sinc((m*pi+g*t)/X);
            end
        end
    end
    %where v is epsilon
    v=0.1;
    RI= R+eye(M)*v;
    y=inv(RI)*QX;
    %Creates list of Q sum of all Yk for each t
    Q=zeros(length(ts)+14,1);
    for t=q:1:(s+9)
        for k=-N:1:N
            Q(t-ii*5)=Q(t-ii*5)+y(k+N+1)*(g/pi)*sinc((k*pi+g*t)/X);
        end
    end
    %Calulating residual
    t=q:1:s+9;
    BLR=0;
    D=zeros(length(Q),1);       
    D=Q(:);
    %for i=1:1:length(Q)
         %for j=1:1:12
             %if zm(i)==j
              %  D(i)=D(i)+S(j);
             %       end
 
        % end
    %end
    for i=1:1:91
        BLR=BLR+sqrt((z(i+5*ii)-(D(i)+MEAN2))^2);
    end
    MV((1+5*ii):(91+5*ii))=MV((1+5*ii):(91+5*ii))+MEAN2;
    BLR
    BLR/90
    D(1:91)=D(1:91)+MEAN2;
    MAA=mean(D(1:91));
    D(92:101)=D(92:101)+MEAN;
    V(q:q+9)=D(92:101);
    s=s+5;
    q=q+5;
    time=time+1
    BLR=0;
end
BLRE=0;
 
%% Residual calculation
for i=1:1:162
    BLRE=BLRE+abs((z(i+91)-(V(i)))^2);
end
LERR=0;
for i=1:1:162
    LERR=LERR+abs((z(i+91)-(LE(i)))^2);
end
length(V);
BLRE
BLRE/length(V(1:162))
LERR
LERR/length(V(1:162))
stots
stots/length(sD(1:162))
length(LE);
length(V);
length(z);
length(MV);
length(sD);
tt=91:1:253;
length(tt)
 
hold on
plot(1:1:90,z(1:90),'k.:',1:1:90,MV(1:90),'r','HandleVisibility','off')
plot(tt,z(91:253),'k.:',tt,V(1:163),'b--',tt,MV(91:253),'r',tt,sD(1:163),'c*')
title('Left band limited casual extrapolation  and salsa extrapolation of ASX stock price,
 5 day forecast for 160 days')
xlabel('Days')
ylabel('Max price $')
line([90,90],[40,100],'Color',[0,0,0],'HandleVisibility','off')
hold all
plot(91:1:253,LE(1:163),'g--')
legend('Raw data','Band limited extrapolation','Five point moving average',
'Salsa Extraploation','Linear extrapolation')
\end{verbatim}
\subsection{"soft" Code created by Ivan Selesnick}
\begin{verbatim}
function y = soft(x, T)
% y = soft(x, T)
%
% SOFT THRESHOLDING
% for real or complex data.
%
% INPUT
%   x : data (scalar or multidimensional array)
%   T : threshold (scalar or multidimensional array)
%
% OUTPUT
%   y : output of soft thresholding
%
% If x and T are both multidimensional, then they must be of the same size.
 
y = max(1 - T./abs(x), 0) .* x;
 
% Ivan Selesnick
% NYU-Poly
% selesi@poly.edu
\end{verbatim}
\end{document}